\documentclass[sigconf]{acmart} 

\AtBeginDocument{%
  \providecommand\BibTeX{{%
    \normalfont B\kern-0.5em{\scshape i\kern-0.25em b}\kern-0.8em\TeX}}}

\acmConference[VEM'2021]{9th Workshop on Software Visualization, Evolution and Maintenance}{September, 2021}{Virtual}

\setcopyright{none} 
\renewcommand\footnotetextcopyrightpermission[1]{} 

\begin{document}



\title{Readability and Understandability Scores for Snippet Assessment: an Exploratory Study}



\author{Carlos Eduardo C. Dantas}
\email{carloseduardodantas@iftm.edu.br}
\affiliation{%
  \institution{Federal University of Uberlândia}
  \country{Brazil}
}

\author{Marcelo A. Maia}
\email{marcelo.maia@ufu.br}
\affiliation{%
  \institution{Federal University of Uberlândia}
  \country{Brazil}
}  

\renewcommand{\shortauthors}{Carlos Eduardo C. Dantas and Marcelo A. Maia.}

\begin{abstract}
Code search engines usually use readability feature to rank code snippets.  There are several metrics to calculate this feature, but developers may have different perceptions about readability. Correlation between readability and understandability features has already been proposed, i.e., developers need to read and comprehend the code snippet syntax, but also understand the semantics. This work investigate scores for understandability and readability features, under the perspective of the possible subjective perception of code snippet comprehension. We find that code snippets with higher readability score has better comprehension than lower ones. The understandability score presents better comprehension in specific situations, e.g. \textit{nested loops} or \textit{if-else chains}. The developers also mentioned writability aspects as the principal characteristic to evaluate code snippets comprehension. These results provide insights for future works in code comprehension score optimization.

\end{abstract}



\maketitle

\section{Introduction}

Code snippets (or code examples) are some lines of  source code that can be reused to show how the developer can solve a specific programming task \cite{Keivanloo2014}. Developers often search for good reusable code snippets on the web \cite{Xia2017}. In average, developers spend 70\% of their time reading programs \cite{Minelli2015}. Some code search engines usually use readability metrics \cite{Scalabrino2018} \cite{Posnett2011} \cite{Buse2010} trying to improve the code snippets ranking \cite{Hora2021APISonarMA} \cite{Moreno2015}. 
These metrics have been employed in recent research, for instance to recommend readable APIs in code snippets \cite{Hora2021APISonarMA} or to evaluate readability changes in projects history \cite{Piantadosi2020}.

However, developers could have subjective perceptions of what means a readable code snippet. The readability metrics are often evaluated with personal opinions as response variable \cite{Oliveira2020}. Consequently, these metrics could produce  false positives/negatives. A potential opportunity to mitigate these mismatches in perception would be combining readability with other related features. Developers need to read and comprehend the code snippet syntax, but also need to understand the code snippet semantic, e.g., the \textit{statements}, \textit{beacons} or \textit{motifs} \cite{Scalabrino2017}. If a  source code is difficult to read, it is also difficult to understand \cite{Boehm1976}. Some metrics have also been proposed to calculate source code understandability \cite{Lin2006} \cite{Lin2008} \cite{Campbell2018}. For instance, the cognitive complexity metric of \textit{SonarSource} tool\footnote{http://apisonar.com/} is related with some aspects of understandability \cite{Marvin2020}.

The main goal in this research is to investigate readability and understandability metric scores on code snippets, to verify their usability on  code snippet comprehension assessment.

We organize the investigation with the following research questions: 

\textbf{RQ \#1)} To what extent the readability and understandability metric scores can be used to code snippet assessment?

\textbf{RQ \#2)} Which characteristics are important to developers on code snippets comprehension evaluation?

To evaluate the metric scores, we asked for five senior developers experienced in approve \textit{pull requests} on \textit{git} repositories (i.e., read, understand and evaluate source code produced by other developers) written in Java language, to evaluate the comprehension of two code snippets extracted from Google and CROKAGE for 30 input queries. A final open question was proposed for them to answer about the relevant code snippets characteristics in their evaluation.

The paper is organized as follows. Section 2 shows a motivating example. Section 3 discusses the related literature. Section 4 presents the study design proposed to collect data, modeling and data analysis approach. The results are reported and discussed in Section 5. Section 6 presents the
threats that could affect the validity of this study. Finally, Section 7  summarizes our observations in lessons learned, and outlines directions for future work.

\section{Motivating Example}

\begin{figure*}[]
\centerline{\includegraphics[width=0.85\textwidth]{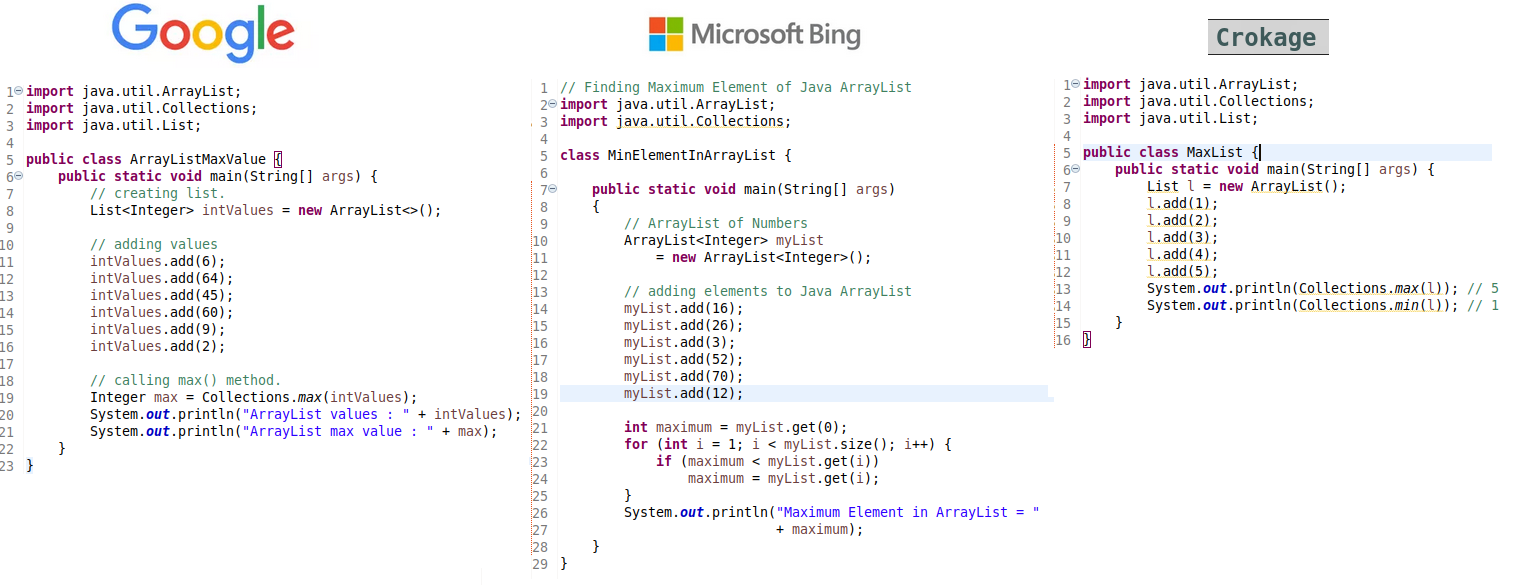}}
\vspace{-4mm}
\caption{Google, Microsoft Bing and Crokage code snippets returned for the input query \textit{Find maximum element of ArrayList in Java} }
\label{fig:comparativo}
\end{figure*}

A motivating example of subjective perceptions is shown in Figure \ref{fig:comparativo}. This example has the first code snippet suggested by Google, Microsoft Bing and CROKAGE \footnote{http://isel.ufu.br:9000/} (tool that provides code snippets and their correspond comprehensive solution for each input query, both mined from Stack Overflow \cite{DBLP:journals/ese/SilvaRRSPDM20}) for the input query \textit{Find maximum element of ArrayList in Java}. The Table \ref{tab:soresultsquery} shows the readability \cite{Scalabrino2018} 
and understandability \cite{Campbell2018} 
score for each suggested code snippet. The CROKAGE code snippet has the lowest \textit{LOC (lines of code}), but Google code snippet has more comments, and the \textit{Collections.max()} method is implemented in a separate line. In CROKAGE, the line 13 has two concepts in the same line, which decreases the readability score in the used metric. The Microsoft Bing code snippet has highest \textit{LOC}, and contains \textit{for loop} statement instead the \textit{Collections.max()} API call. In readability score, the CROKAGE code snippet is nearest to Microsoft Bing. But in understandability score, CROKAGE and Google has the same value, both better than Microsoft Bing code snippet.

The  example shows the readability score could have divergent opinions, because some developers could prefer the lowest \textit{LOC} instead of one concept per line. In this example, the understandability score has the trade-off between internal API call or a \textit{for-loop} statement, which could be less divergent than readability metric because generally snippets using internal APIs has less complexity and are reusable in other programs \cite{Moreno2015}. The readability and understandability metrics could complement each other on code comprehension.

\begin{table}[h]
\centering
\caption{Readability and understandability (higher is better) scores for Google, Microsoft Bing and CROKAGE code snippets in Figure \ref{fig:comparativo}}
\vspace{-4mm}
\label{tab:soresultsquery}
\begin{tabular}{|l|c|c|} 
\hline & \multicolumn{2}{c|}{Score} \\
\cline{2-3} \raisebox{1.2ex}{Web Search Engine} & Readability & Understandability  \\
\hline
 Google    & 0.67 & 1.0 \\ \hline
 Microsoft Bing    & 0.51 & 0.8 \\ \hline
 CROKAGE    & 0.55 & 1.0 \\
 \hline
\end{tabular}
\vspace{-4mm}
\end{table}

\section{Related Work}

Several code search engines has been purposed to rank code snippets using the readability feature as part of the overall score. \textit{Hora} \cite{Hora2021Google} investigated how Google, a general-purpose web search engine rank the code snippets in terms of readability and reusability features. Their findings shows that readable and reusable code snippets are not necessarily top ranked, but other aspects as didactic code snippets or pages with multiple code snippets are more likely to be top ranked. Our research is not interested in discover how Google rank their code snippets, but to provide insights if readable and understandable code snippets are relevant for developers, and then other future researches could use these features as a  part of overall score on new code search engines.   

In other research, \textit{Hora} \cite{Hora2021APISonarMA} constructed the API \textit{Sonar tool}, mining code snippets from 100 Java APIs on  \textit{github} to generate collections of API code snippets. He is also using readability to top rank readable API code snippets. The insights in our research could be useful to provide a better ranking, considering understandability in certain situations. \textit{Moreno et al.} develop the \textit{Muse} approach to rank code snippets producing an overall score using readability and reusability feature. But this research has employed other readability approach \cite{Buse2010}, and the other mentioned researches has used the \textit{Scalabrino et al.} readability approach \cite{Scalabrino2017}.

Another features could be used to produce an overall score. \textit{Oliveira et al.} \cite{Oliveira2020} introduced a separation between readability and legibility features, where legibility is related to how easy to identify elements in a program. For example, code without indentation or more than one statement in the same line contributes to decrease legibility. The readability tool used in this research consider some legibility aspects on their metrics, e.g., one concept per line.

Some researches studied the correlation between readability and understandability features. \textit{Boehm et al.} \cite{Boehm1976} pointed the source code readability is related to its respective complexity and understandability, i.e., if the source code is difficult to read, it is also difficult to understand. But even \textit{easy-to-read} source code can be difficult to understand, as presented by \textit{Scalabrino et al.} \cite{Scalabrino2017}. Therefore, readability and understandability are employed as different features: while readability measures the effort to understand a code snippet in syntactic aspect, understandability measures complexity in dynamic aspect \cite {Posnett2011}, i.e., both metrics complement each other in measure code comprehension.

The understandability feature has divergent results about metrics. \textit{Scalabrino et al.} \cite{Scalabrino2017} made a study with 121 distinct metrics and found that none of them is relevant to measure understandability. However, \textit{Marvin et al.} \cite{Marvin2020} found correlations between cognitive complexity metric \cite{Campbell2018} with subjective ratings of understandability, which is a relatively positive insight about the effectiveness of this metric. The cognitive complexity proposal is similar to the cyclomatic complexity of \textit{McCabe} \cite{McCabe1976}. However, cognitive complexity aims to mitigate the limitations of cyclomatic complexity, such as source code nesting problem \cite{Suleman2013}, and address modern language structures such as \textit{try/catch} or \textit{lambdas}. 

\section{Study Design}

\begin{figure}[h]
\centerline{\includegraphics[width=0.3\textwidth]{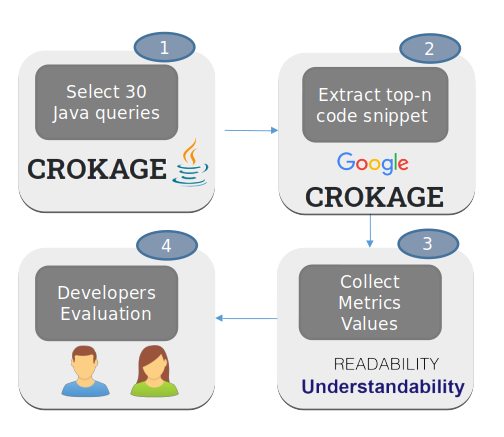}}
\vspace{-4mm}
\caption{Overall architecture proposed in this research}
\label{fig:architeture}
\end{figure}

The Figure \ref{fig:architeture} shows the overall approach to answer the two research questions. The major steps are: (1) Select Input Queries, (2) Extract Code Snippets, (3) Collect Metrics Values and (4) Developers Evaluation. The details of each step is in the following subsections. A replication package, including the readability and understandability tools, as the questions, code snippets characteristics, evaluations and the  instructions for  reproduction is available \cite{carlos_eduardo_c_dantas_2021_5224346}.  

\vspace{-4mm}

\subsection{Select Input Queries}

CROKAGE have already collected 10,370 Java programming input queries performed from developers of more than 80 countries around the world. These queries contains users searches for code snippets implementing specific programming tasks in Java language. We selected 30 popular input queries performed by different users (i.e., distinct \textit{IP address}) to conduct the experiment. To find the popular queries, we processed each query removing all punctuation symbols, stop words\footnote{https://bit.ly/1Nt4eMh} and small queries (i.e., size lower than three). And then, we ordered the queries by the number of occurrences. 

\subsection{Extract Code Snippets}

This step consists in extract two code snippets to each input query. These code snippets were extracted manually from Google (using private browsing on \text{Google Chrome} to avoid caches, user preferences) and CROKAGE. We collected the first code snippet that match each input query (e.g. assign the input query \textit{"how to sort an array in java?"} has code snippets related to sort arrays and not \textit{java.util.Set} or any other data structure), and contains reusable source code (e.g., we discarded code snippets  
containing references to unknown methods or variables) \footnote{https://tinyurl.com/jtjh75bx}. For each query, we added the tokens \textit{"in java"}, to receive Java code snippets in the top ranking Google recommendations.

\begin{figure*}[!h]
\centerline{\includegraphics[width=1.0\textwidth]{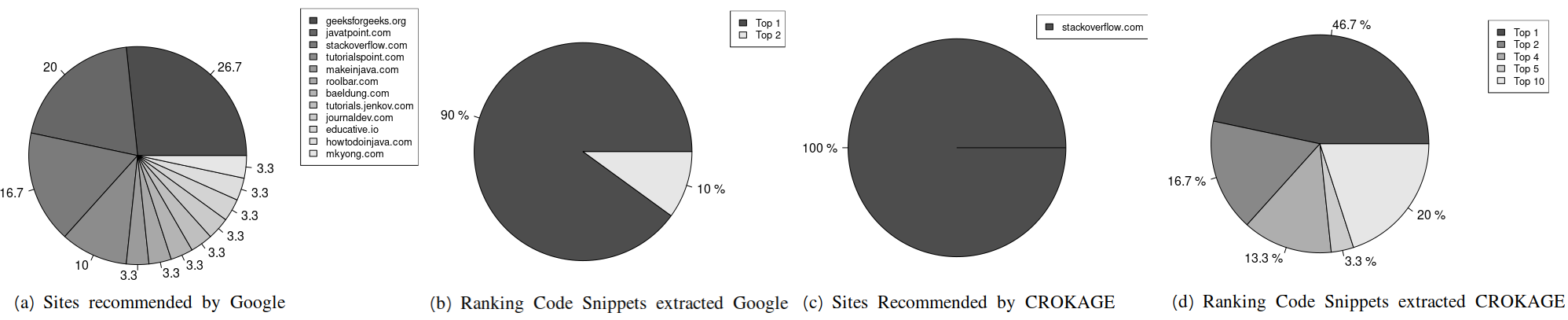}}
\vspace{-4mm}
\caption{Sites and ranking of code snippets returned by Google and CROKAGE}
\label{fig:extract}
\end{figure*}

The Figure \ref{fig:extract} shows the sites where code snippets were extracted, and their ranking in the web search. Using Google as web search, most of the snippets were in first result (top-1) and extracted in \textit{geeksforgeeks.org}, \textit{javatpoint.com} and \textit{stackoverflow.com} sites. Using CROKAGE, all code snippets were extracted from \textit{stackoverflow.com}, and 46.7\% were the first result.

\subsection{Collect Metrics Values}

This step consists in extract the readability and understandability scores of each code snippet. 

To measure readability, we use the prediction model proposed by \textit{Scalabrino et al.} \cite{Scalabrino2018}, which was used in other recent researches \cite{Hora2021APISonarMA} \cite{Hora2021Google} \cite{Piantadosi2020}. This model\footnote{https://dibt.unimol.it/report/readability/} includes a set of metrics including comments, identifiers consistency, textual coherence and number of meanings. The model produces scores between 0 (low readability) and 1 (high readability). 

To measure understandability, we use the cognitive complexity code-based metric proposed by \textit{Campbell} \cite{Campbell2018} and available in \textit{SonarSource} tool. This metric were evaluated and employed in some past reaseaches \cite{Marvin2020} \cite{Wyrich2021}. To measure understandability, we propose an adoption metric as follows:

\vspace{-4mm}
\[ understandability(cs_i) =
  \begin{cases}
    1 - \frac{\#cc}{\#mcc}    & \quad \text{if } \#cc < \text{15}\\
    0.0 & \quad \text{otherwise}
  \end{cases}
\]
\vspace{-3mm}

\textit{\#cc} is the complexity cognitive score extracted from \textit{SonarSource} tool for the code snippet \textit{$cs_i$}. The \textit{\#mcc} is the maximum recommend complexity cognitive value \footnote{https://tinyurl.com/zfka2ew2} 
, \textit{\#mcc} = 15. If a code snippet reaches \textit{\#cc} >= 15, the score output will be 0. This metric produces scores between 0 (low understandability) and 1 (high understandability). 

\subsection{Developers Evaluation}

We invited five senior developers to analyse the comprehension of 60 code snippets. These developers are (5+ years) experienced as team leaders on Java projects, and they often evaluate \textit{pull requests} submitted from other developers on their teams. \textit{Pull requests} are usually rejected because could have issues in code comprehension or understanding \cite{Papadakis2020}. 

For each of the 30 queries, the code snippets of each solution (Google and CROKAGE) were presented side by side, and asked for the developers to provide a \textit{likert} value from 1 to 5 for the comprehension of each suggested code snippet. The code snippets for 30 questions are distributed on the following criteria:

\begin{enumerate}
    \item 10 questions with higher readability in one code snippet and similar understandability in both code snippets.
    \item 10 questions with higher understandability in one code snippet and similar readability in both code snippets.
    \item 10 questions with higher readability and understandability in one code snippet compared to the other.
\end{enumerate}

The objective is to obtain the developers comprehension evaluation on each feature isolated, and both combined, to evaluate if the code snippets with better readability and understandability score are significant better evaluated than the lower ones. And finally, the developers answered in a open question the characteristics they included to evaluate each code snippet comprehension. 

\section{Results and Findings}

In this section, the results will be shown according to each research question. This research used Krippendorff's $\alpha$ reliability  coefficient \cite{Krippendorff2011ComputingKA} to verify the agreement between the five developers. We obtained $\alpha = 0.334$, which is a low agreement. Only six of 30 code snippets with higher score had perfect agreement between the developers, as the same with one of 30 code snippets with lower score. The low agreement implies a subjective source code comprehension analysis.





\textbf{RQ \#1) To what extent the readability and understandability metric scores can be used to code snippet assessment}

\begin{figure*}[!h]
\centerline{\includegraphics[width=1.0\textwidth]{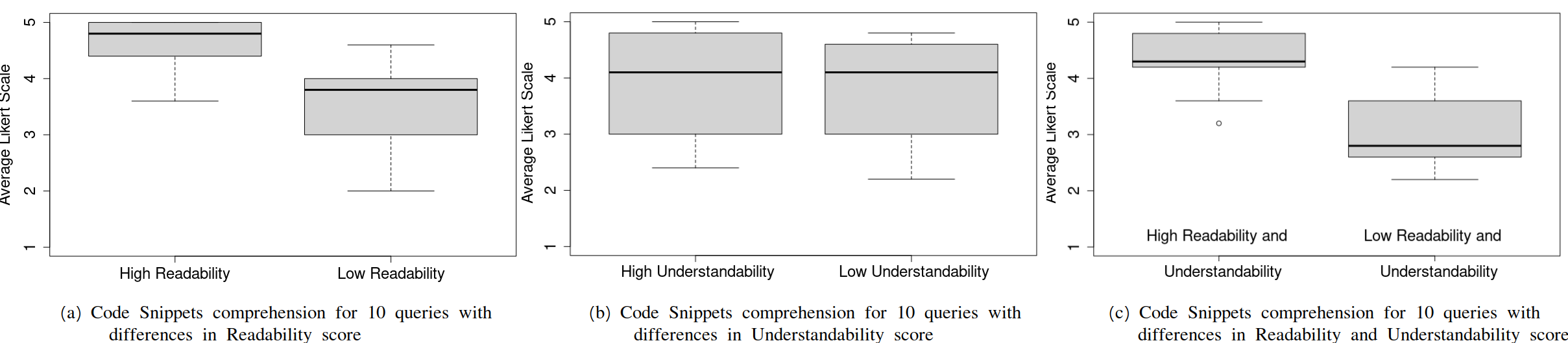}}
\caption{Box plots of Code Snippets comprehension}
\vspace{-5mm}
\label{fig:boxplot}
\end{figure*}

The Table \ref{tab:features} shows the likert evaluation results. We run Wilcoxon signed-rank test using a confidence level of 99\% (p-value<0.01), and the comprehension of code snippets with higher readability score are statistically better than snippets with lower readability score. The Figures \ref{fig:boxplot}a and \ref{fig:boxplot}c shows a better rate for code snippets with higher readability score. This result indicates the readability score match the developers perception on code comprehension. The Figure \ref{fig:boxplot}b shows statistically the same rate for higher and lower understandability score (as shown in Table \ref{tab:features}), i.e., in this analysis, the understandability score is not relevant for the five developers on code snippets comprehension. 

We investigate the understandability likert evaluation, and more significant likert differences were found in specific situations. For example,  some code snippets with higher understandability code uses internal Java API methods to implement the task, and the code snippet with lower understandability score is using \textit{nested loops} or \textit{if-else chains}. However, the developers did not pointed difficult to comprehend few addictions of \textit{loop}, \textit{condition} or \textit{try/catch block} on code snippets with lower understandability score. We extracted the highest understandability score differences between code snippets, and in this scenario, the effect size increases to 0.812 with \textit{p-value = 0.06}, i.e., the understandability feature has better effect size in code snippets with high differences on understandability score. For example, on the input query \textit{How to remove an element from an array in Java?}, the average likert for a code snippet with higher understandability score, using the \textit{java.util.Stream.filter()} solution and  \textit{LOC} = 51 is 4.2, and the code snippet with \textit{for nested loop} solution and \textit{LOC} = 25 is 3.2. In another example, in the query \textit{How to split a string in Java?}, the code snippet with lower understandability score has four extra lines with an additional \textit{for statement} to print each splitted string. But the developers evaluate this code snippet as the best comprehend solution, i.e., the additional \textit{for statement} had positive influence on their evaluation.


\begin{table}[]
\centering
\caption{P-value and effect-size values for each feature}
\vspace{-4mm}
\label{tab:features}
\begin{tabular}{|l|c|c|} 
\hline Feature & p-value & effect-size  \\
\hline
 Readability    & < 0.01 & 0.857 \\ 
 Understandability & 0.91 & -  \\
 Readability + Understandability & < 0.01 & 0.838 \\ \hline
\end{tabular}
\vspace{-4mm}
\end{table}

\noindent
\begin{center}
\fbox{\begin{minipage}{25em}
\textbf{RQ \#1 Answer:} The readability score is associated with code snippets comprehension, but the low agreement between developers suggests subjective perception. The understandability score is related to code comprehension in specific situations, e.g., \textit{nested loops} or \textit{if/else chains}. These specific situations has higher difference on code snippets understandability score. 
\end{minipage}}
\end{center}
\vspace{.4em}

\textbf{RQ \#2) Which characteristics are important to developers on code snippets comprehension evaluation?}

The Table \ref{tab:characteristics} shows the mentioned characteristics on code snippets evaluation. Four of five developers (80\%) mentioned writability aspects, (e.g., simplicity and clarity on write code, expressiveness on self documented source code avoiding comment lines). Three of five developers (60\%) mentioned about aspects of variables and methods (e.g., camel case pattern, easy naming comprehension, variable declarations aspects as default values). Also three of five developers (60\%) mentioned comments, but all of them mentions that comments are only useful for source code with higher complexity. In their opinion, comments increases the number of lines, and may not be useful for code comprehension, giving a larger extension than the necessary. Other three of five developers (60\%) mentioned complexity. Some \textit{nested for loops} and \textit{if else chains} were used in some code snippets instead of simple Java internal API calls. The reusability aspect was mentioned, i.e., more complex code snippets is more difficult to reuse in other software development project. One developer mentioned \textit{LOC}, because in his opinion, fewer lines written in Java is generally easier to comprehend. Finally, one developer mentioned performance, i.e., if the solution is appropriate to run in production environment.

\begin{table}[]
\centering
\caption{Characteristics mentioned by the developers on code snippets evaluation}
\vspace{-4mm}
\label{tab:characteristics}
\begin{tabular}{|l|c|} 
\hline Characteristic & \% mentions  \\
\hline
 Writability    & 80\% \\ 
 Variable and method aspects & 60\% \\ 
 Comments    & 60\%  \\ 
 Complexity & 60\% \\ 
 Lines of Code (LOC) & 20\% \\
 Performance & 20\% \\ 
 \hline
\end{tabular}
\vspace{-4mm}
\end{table}

\noindent
\begin{center}
\fbox{\begin{minipage}{25em}
\textbf{RQ \#2 Answer:} Some characteristics mentioned by developers are related to readability feature (variable and method aspects, comments, \textit{LOC}) and understandability feature (complexity). However, most of the developers mentioned aspects of writability, which opens for new approaches investigating metrics for this characteristic.
\end{minipage}}
\end{center}
\vspace{.4em}

\section{Threats to Validity}

The main threat in this research is related to study generalization (programming language, number of participants, number of queries and code snippets).

\textbf{Programming Language}: the results are restricted to Java programming language, specially because limitations of the queries and the readability tool. The cognitive complexity tool supports more programming languages, e.g., Python, Javascript.

\textbf{Number of participants}: the five senior developers work in different companies (i.e., different core business and applications). We try to mitigate the few number of developers selecting team leaders with experience in evaluate and approve \textit{pull requests} written by other developers. But novice developers could have different perceptions about readable and understandable code snippets.

\textbf{Number of queries and code snippets}: this research extracted 60 code snippets for 30 queries. A higher number of queries is an important factor to future research (e.g. produce an unified score using a combination of features). 

\textbf{False positives/negatives}: the readability and understandability metrics could have some false positives. Complexity cognitive metric uses static heuristics. To minimize the effect, we manually analysed the score of each code snippet.

\section{Conclusions and Future Work}

In this exploratory study, a quality analysis is conducted across code snippet comprehension using readability and understandability metric score. Our findings suggest the readability score could be used on code snippet assessment, e.g., code search engines. The understandability score have more subjective perceptions, specially in lower score differences between code snippets.

These results provide insights for several improvements. Future research could propose an empirical study to optimize a new unified score between understandability and readability features. Another code comprehension features could be evaluated, such as legibility and writability. Finally, a study with developers profiles could be addressed. Novice developers would have different perceptions about code comprehension than the team leaders used in this research. 

  \bibliographystyle{ACM-Reference-Format}
  \bibliography{acmart}

\end{document}